# Assessing Mortality of Blunt Trauma with Co-morbidities.

Clive E Neal-Sturgess
Emeritus Professor of Mechanical Engineering
University of Birmingham.

*"..a living organism tends to approach the dangerous state of maximum entropy, which is death." Erwin Schrodinger, What is Life? 1944*.

**Abstract**

**Objectives:** To obtain a better estimate of the mortality of individuals suffering from blunt force trauma, including co-morbidities.

**Methodology:** The Injury severity Score (ISS) is the default world standard for assessing the severity of multiple injuries. ISS is a mathematical fit to empirical field data. It is demonstrated that ISS is proportional to the Gibbs/Shannon Entropy. A new Entropy measure of morbidity from blunt force trauma including co-morbidities is derived based on the von Neumann Entropy, called the Abbreviated Morbidity Scale (AMS).

**Results:** The ISS trauma measure has been applied to a previously published database, and good correlation has been achieved. Here the existing trauma measure is extended to include the co-morbidities of disease by calculating an Abbreviated Morbidity Score (AMS), which encapsulates the disease co-morbidities in a manner analogous to AIS, and on a consistent Entropy base. Applying Entropy measures to multiple injuries, highlights the role of co-morbidities and that the elderly die at much lower levels of injury than the general population, as a consequence of co-morbidities. These considerations lead to questions regarding current new car assessment protocols, and how well they protect the most vulnerable road users.

**Injury Severity Scaling**

Injury scaling, as a means of classifying the severity of impact trauma, has a long history. Some of the earliest research into impact trauma was conducted at Cornell University Medical School in 1952 by De Haven and colleagues (Petrucelli 1993), and was related to aircraft crashes. The sixties saw many developments when a number of first generation methodologies were also proposed by: Robertson et.al. (Robertson, McLean et al. 1966), Nahum et.al. (Nahum, Siegel et al. 1967), Mackay (Mackay 1968), Van Kirk and Lange (Van Kirk and Lange 1968), States and States (States and States 1968), Keggl (Keggl 1969), and Campbell (Cambell 1969).

In 1968 Ryan and Garrett (Ryan and Garrett 1968) revised De Haven's scale, and considered energy dissipation, as well as threat to life, as criteria. The Comprehensive Research Injury Scale (CRIS) was developed using these concepts as shown in Table 1.

|         | Energy Dissipation | Threat to Life |
|---------|-------------------|----------------|
| Level 1 | Little            | None           |
| Level 2 | Minor             | Minor          |
| Level 3 | Moderate          | Moderate       |
| Level 4 | Major             | Severe         |
| Level 5 | Maximum           | Maximum        |

Table 1: Energy Dissipation and Threat to Life

The Abbreviated Injury Scale (AIS) was first officially published in 1972, revised in 1974 and 75 and published in manual format in 1976 (Scaling 1976). Revisions were published in 1980, 85, 90, 98 and 2005 (Gennarelli and Wodzin 2006). AIS is usually described as a non-linear ordinal scale; this is only one interpretation. If the genesis of AIS is followed it is obvious that it is a non-linear integer scale related to energy dissipation. It is integer simply because no fractional AIS measures have been defined, as the clinical resolution would not support such fine scale measures. The AIS score has proven to be the "system of choice" (Petrucelli 1993), and has been documented in many articles (Medicine 1983; Stevenson, Segui-Gomez et al. 2001). A number of user groups have modified the basic AIS scale to account for particular types of injury and harm, and a case has been made for unification(Garthe and Mango 1998).

Baker et.al. (Baker and O'Neil 1974; Baker 1974) found that the AIS score was a non-linear predictor of mortality. This is not a fundamental problem as non-linearities can be easily accommodated. However, it was found that the death-rate of persons with two or more injuries was not simply the sum of the AIS scores. This led to the introduction of the empirical Injury Severity Score (ISS) as a means of approximately linearising the data with regard to the probability of fatality. The ISS is the sum of squares of the maximum AIS code (MAIS) in each of the three most severely injured body regions (Baker 1974).

ISS has been criticised as it only allows the counting of one MAIS value per body region, where in reality there are often more that on MAIS value in a given body region. The New Injury Severity Scale (NISS) was introduced (Stevenson, Segui-Gomez et al. 2001) to allow the counting of more than one MAIS value for a given body region. It was shown that NISS is marginally more effective than ISS. Both ISS and NISS are calculated on the basis of ordered triplets, and so there are a significant number of ISS or NISS values that cannot be achieved in practice.

The fundamental problem with ISS, as recognised by Boyd et.al. (Boyd, Tolson et al. 1987), is that ISS essentially only measures the traumatic insult, and the combined effect of the traumatic insult and the persons underlying medical and physiological reserve is also very important. This led to the ASCOT scoring system, and Champion has shown by using logistic regression that ASCOT has a better predictive capabilities than ISS or NISS (Champion, Copes et al. 1996). This research finally led to the TRISS method (Trauma and Injury Severity Score), which includes the ISS score, the RTS (Revised Trauma Score, and the patient's age) (Boyd, Tolson et al. 1987), which is available as an online calculator (Boyd, Tolson et al. 1987). Age is a very important variable in trauma, as it is the principal proxy for frailty, however both ASCOT and TRISS are very coarse grained with respect to age, leading to misdiagnosis in a number of seriously injured casualties (Demetriades, Chan et al. 1998) – a better model of the effects of co-morbidities is required.

A considerable amount of work has been done on the subject of entropy and ageing, beginning with Schrodinger in 1944 (Schrodinger 1944), and followed by Strehler (Strehler 1960; Strehler 1962). It is a reasonable assumption that entropy increases throughout life, either through genetic mutations, disease, or misrepair (Atlan 1975; Riggs 1993; Silva and Annamalai 2008; Salminen and Kaarniranta 2010), probably in a non-linear manner (Neal-Sturgess 2010). This is shown schematically in Fig.1.

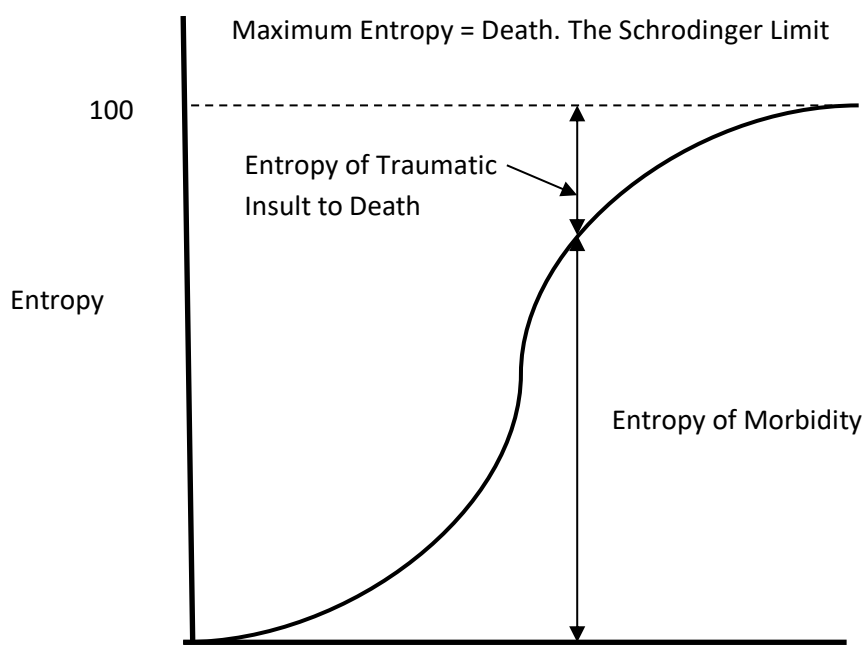

Fig. 1. Schematic of the Entropy of Life Curve.

In a relatively recent large study of extremely injured geriatric casualties (Duvall, Zhu et al. 2015) it was found that ISS plus age and co-morbidities did not correlate well with mortality, and they concluded that physiological information was necessary; although this was not tested. However, the sample was very particular, and the inclusion of co-morbidities was simply how many existed; hence there was no coherent model involved (see later). In a recent paper (Wang, Ye et al. 2017. ) {abstract in English} the authors found that over the period 1990-2013 in China, overall the death rates from motor vehicle injuries increased, although the burden of disease on motor vehicle outcomes decreased. The 15-49 years old age group had the highest burden of disease.

Age, as was said earlier, is a proxy for Frailty. Frailty is a multi-dimensional complex clinical condition, capable of diagnosis, but for which no commonly accepted definition exists (Rockwood, Song et al. 2005). There have been a number of measures of frailty, some functional, and some operational (deVriesa, Staal et al. 2010). However, they frequently are complex to use (can include up to 70 variables) (Rockwood, Song et al. 2005), and time consuming, so have not gained wide acceptance.

All measures of frailty, irrespective of the scoring systems involved, include co-morbidities (Xue 2011), which here are considered important in the context of adverse outcomes from vehicle crashes (Wang, Ye et al. 2017. ), the question is which ones, and how to include in the combined effects of morbidity and trauma. One of the reasons ISS has been so widely adopted is its simplicity and rapidity of calculation, which is the starting point here to include co-morbidities. Also when aggregating multi-dimensional scores it is vital to have a common homogeneous base (commensurable), otherwise it is like adding apples and pears. It is proposed here to use von Neumann Entropy as the common base.

The problem then is to estimate the Entropy of Morbidity, in an earlier paper (Neal-Sturgess 2011)  the author derived the entropy of morbidity in a coarse grained manner, and derived the combined entropy of morbidity and trauma based on the Gibb's Entropy. However, a finer grained estimate of the entropy of morbidity is ideally required. Here it is assumed that the principal cause of adverse outcomes from crashes are pre-existing medical conditions (PMC's). An Abbreviated Morbidity Score (AMS), analogous to the Abbreviated Injury Score (AIS) is postulated, and a Combined Mortality Score (CMS) is calculated based on summing the von Neumann's entropy, which is a generalisation of the Gibbs/Shannon entropy (Petz 2001.).

**Theory:**

When examining the ISS it is immediately obvious that it preferentially weights the higher AIS scores, but this does not explain how it works. A second consideration is that it has the form of a vector resultant.

If a human body is subject to multiple injuries to a given body region AIS, then the injuries to a given body region (i = 1-3) can be expressed as state vectors in a Hilbert "Injury" Space, and if it is diagonalized using the maximum AIS i.e. MAIS.  The vector resultant then becomes ISS as:

$$ISS = MAIS_1^2 + MAIS_2^2 + MAIS_3^2 \qquad (1)$$

This can be extended to multiple dimensions (multiple body regions).

As MAIS is proportional to the probability of death (p), and the relationship is non-linear, then the vector resultant is also a vector of probabilities.

Similarly, if a human body is subject to multiple injuries to a given body region $AIS_{i-n}$ which are represented as the Eigenvectors (rows) of an Injury Density Matrix IDM, and if the maximum AIS (MAIS) is treated as the Eigenvalue of the injury Eigenvector (diagonalisation), an injury density matrix I can be constructed as shown:

$$IDM = \begin{bmatrix} AIS_{11} & AIS_{12} & AIS_{13} \dots \dots \\ AIS_{21} & AIS_{22} & AIS_{23} \dots \dots \\ AIS_{31} & AIS_{32} & AIS_{33} \dots \dots \end{bmatrix} \qquad (2)$$

Where $AIS_{11}$, $AIS_{22}$, and $AIS_{33}$ etc. are the Eigenvalues of the Eigenvector, and are the MAIS's.  The associated Eigenvector in this case represents the body region injured.

There is no need to conduct a principal Component Analysis (PCA), as the AIS vectors are the Eigenvectors

Then the 2nd power of the Trace is:

$$Tr\,(I)^2 = \text{AIS}_{11}\text{AIS}_{11} + \text{AIS}_{22}\text{AIS}_{22} + \text{AIS}_{33}\text{AIS}_{33}\ldots = \sum_{i=1}^{n}\sum_{i=1}^{n} a_{ij}a_{ji} \qquad (3)$$

Which is the Injury Severity Score, ISS.

If the MAIS's are replaced by the respective probabilities of death ($p_{ij}$) in the form of the Gibbs Entropy i.e. ($\rho log \rho$) then they can also be regarded as a density matrix, and the relevant value is the von Neumann entropy as:

$$S(\rho) = -kTr(\rho log \rho) \qquad (4)$$

which is a generalisation of both the Gibbs and Shannon Entropies (Li and Busch 2013).

To include morbidities, if a global coordinate system of mortality is constructed, with the co-morbidities and trauma as sub spaces, then for the co-morbidities as the Abbreviated Morbidity Score {AMS(n)}, it is possible to construct a Morbidity Density Matrix (MDM) as;

$$MDM = \begin{bmatrix} AMS_{11} & AMS_{12} & AMS_{13}\ldots\ldots \\ AMS_{21} & AMS_{22} & AMS_{23}\ldots\ldots \\ AMS_{31} & AMS_{32} & AMS_{33}\ldots\ldots \end{bmatrix} \qquad (5)$$

Where the $AMS_{11}$, $AMS_{22}$, and $AMS_{33}$, are the maximum AMS's i.e. MAMS, then a vector resultant **R** can be constructed for the co-morbidities, and a vector resultant of **r**, for the trauma as:

$$\boldsymbol{R^2 = MAMS_1^2 + MAMS_2^2 + MAMS_3^2} \qquad \textbf{(6)}$$

$$\boldsymbol{r^2 = MAIS_1^2 + MAIS_2^2 + MAIS_3^2} \qquad (7)$$

Then if CMS is the combined mortality score,

$$\boldsymbol{CMS = \lambda R^2 + r^2} \qquad \textbf{(8)}$$

Where λ is a scalar, determined empirically.

**Discussion:**

It has been shown in an earlier paper that $r^2$ is proportional to ISS (Neal-Sturgess 2011), with good correlation; so here attention is focussed on the pre-existing medical conditions or co-morbidities. The next question is what co-morbidities to include, here the Leading Causes of Death are taken as a reasonable starting point from which to rank the importance of co-morbidities.

The leading causes of Death, excluding accidents and suicides are, with a TENTATIVE value for AMS (2017):

| Cause of Death | % of Total | Probability of death | TENTATIVE AMS |
|---|---|---|---|
| Heart Disease | 23.4 | 0.234 | 5 |
| Cancer | 22.5 | 0.225 | 5 |
| Respiratory Disease | 5.6 | 0.056 | 4 |
| Stroke | 5.1 | 0.051 | 4 |
| Diabetes | 2.9 | 0.029 | 3 |
| Pneumonia | 2.1 | 0.021 | 3 |
| Kidney Disease | 1.8 | 0.08 | 3 |

Table 1.

It is probably not profitable to include causes of death lower than the ones quoted, except as a score of 2 or 1, as this would only have only a very marginal effect on the result.

As it is assumed here that the principal components of frailty are the pre-existing medical conditions (PMC's), of which age is used as a crude proxy, then from Fig 2, it can be seen that for 25 < ISS <45 the curves for the 0-49 age group and the 70+ group are approximately parallel, and separated by around 35 points. From the table above, with tentative AMS values, it is considered that the 70+ age group would most likely suffer from heart disease and possibly cancer, plus other conditions,; this would give an ISS uplift of at least $5^2 + 5^2 = 50$. However, in the round, casualties would probably suffer from one condition at score 5, plus others. Therefore, from equation 8, it would appear that a first approximation $\lambda \approx 1$.

This example may not be accurate, as it is based on simplified assumptions, however the trend is in the right direction. Although the mathematics may seem complex, the final methodology is very simple to apply; in fact the AMS values are easily committed to memory.

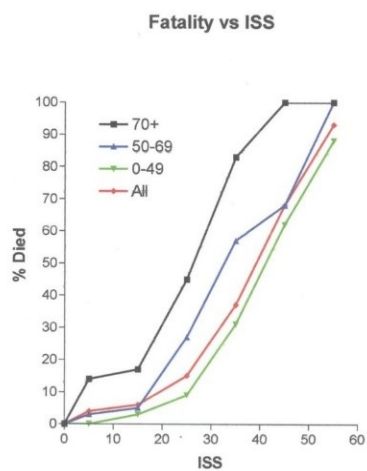

Fig. 2.

One of the problems with the approach outlined is that a co-morbidity is either on or off (a binary approach), whereas the scalar λ can be used to fine tune the system. The task is to scale $R^2$ for morbidity i.e. determine AMS's and λ, this will require large database study, and in the UK's case, access to the NHS Summary Database. In a manner analogous to AIS, a panel of trauma surgeons is needed to produce estimates of AMS's, and it is possible that the most highly rated diseases in mortality are not the only ones which interact with trauma to give adverse outcomes, such as osteoporosis.

**Conclusions:**

A new Entropy measure of mortality from blunt force trauma additionally including co-morbidities of disease is derived based on the von Neumann Entropy. The trauma measure has been applied to a previously published database, and good correlation is achieved. Here the existing trauma measure is extended to include the co-morbidities of disease by calculating an Abbreviated Morbidity Score (AMS), which encapsulates the disease co-morbidities in a manner analogous to AIS; the resulting methodology is simple to apply.

Applying Entropy measures to multiple injuries, highlights the role of co-morbidities, and that the elderly die at much lower levels of injury than the general population. These considerations lead to questions regarding current new car assessment protocols, and how well they protect the most vulnerable road users.